\documentclass[prb,reprint]{revtex4-1} 
\usepackage{subcaption}
\usepackage[thinc]{esdiff} 
\usepackage{empheq} 
\usepackage{amsmath}
\usepackage{amsfonts}
\usepackage{hyperref}
\usepackage{color}

\definecolor{myboxcolor}{rgb}{0.96, 0.96, 0.96}

\usepackage{xspace}
\newcommand\thedata {$\{(t_i,h_{\text{obs}, i})\}_{i=1}^{N}$\xspace}
\newcommand\thedatanomath {\{(t_i,h_{\text{obs}, i})\}_{i=1}^{N}}
\newcommand\themodel {$h(t; h_0, \boldsymbol \alpha, \boldsymbol\theta)$\xspace}
\newcommand\themodelnomath {h(t; h_0, \boldsymbol \alpha, \boldsymbol\theta)}
\newcommand\thevars{h_0, \boldsymbol \alpha, \boldsymbol \theta, \sigma^2}

\definecolor{shadecolor}{rgb}{0.97, 0.97, 1.0}
\definecolor{titlecolor}{rgb}{0.96, 1.0, 0.98}
\newsavebox{\mysaveboxM} 
\newsavebox{\mysaveboxT} 
\newcommand*\Garybox[2][Example]{
	\sbox{\mysaveboxM}{#2}%
	\sbox{\mysaveboxT}{\fcolorbox{black}{titlecolor}{#1}}%
	\sbox{\mysaveboxM}{%
	\fcolorbox{black}{shadecolor}{%
	\makebox[\linewidth-2em]{\usebox{\mysaveboxM}}%
}%
}%
\usebox{\mysaveboxM}%
\makebox[0pt][r]{%
\makebox[\wd\mysaveboxM][c]{%
\raisebox{\ht\mysaveboxM-0.5\ht\mysaveboxT
+1.6\dp\mysaveboxT-0.5\fboxrule}{\usebox{\mysaveboxT}}%
}%
}%
}

\usepackage{tcolorbox} 
\tcbuselibrary{breakable}
\newtcolorbox[auto counter
]{mytcbox}[2][]{
title=#2, #1,
colback=white,
colframe=black,
fonttitle=\bfseries,
parbox=false
}

\begin{document}

\title{Inferring the shape of a solid inside a draining tank from its liquid level dynamics}

\author{Gbenga Fabusola}
\author{Cory M. Simon}
\email{cory.simon@oregonstate.edu}
\affiliation{School of Chemical, Biological, and Environmental Engineering. Oregon State University. Corvallis, OR, USA.}





\begin{abstract}
We aim to reconstruct the shape of an exogenous, heavy solid contained in a tank from measurements of the liquid level in the tank as it drains (driven by gravity) through a small orifice in its side. (Because the solid displaces liquid, the rate of decrease of the liquid level provides information about the cross-sectional area of the solid at that height; as the liquid level drops, it ``scans’' the area of the solid as a function of height.) We combine mathematical modeling, Bayesian statistical inversion, Monte Carlo simulation, and wet experiments of a tank draining of water to demonstrate and test our ability to infer the cross-sectional area of the exogenous solid as a function of height. In our experiment, the posterior distribution over the [held-out] shape of the solid (a bottle) agreed reasonably well with our length-measurements ($<$10\% mean reconstruction error on its radius). Our approach may be practically useful to non-destructively characterize the geometry of an unknown solid, or a packed bed of solid particles, contained in an opaque tank.

\begin{center}
	\includegraphics[height=0.2\textwidth]{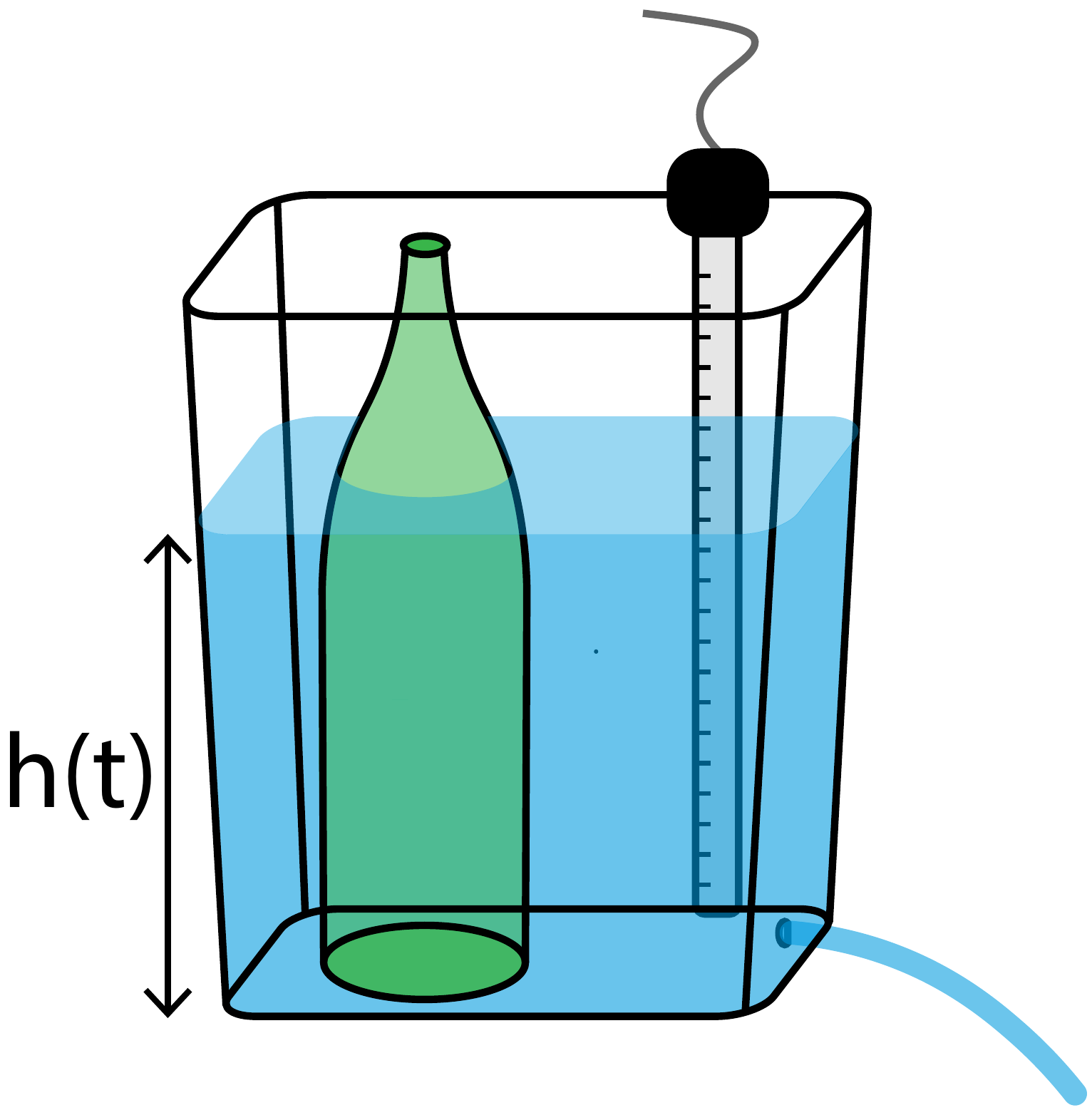}
	\includegraphics[height=0.2\textwidth]{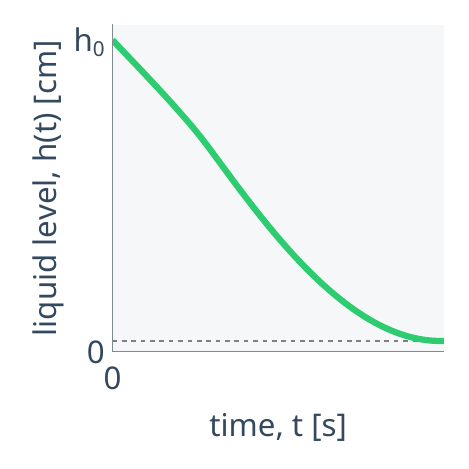}
\end{center}

\end{abstract}

\maketitle 

\section{Introduction}
Throughout engineering and the applied sciences, we may encounter an open-top tank, draining of liquid via gravity-driven flow through a small orifice (a \emph{draining tank}).
Mathematical models of the dynamics of the liquid level in a draining tank are useful for designing the tank and orifice geometry, predicting emptying times, forecasting outlet flow rates, controlling the liquid level by manipulating an input stream, and inferring the liquid level from the outlet flow rate \cite{d2021torricelli,seborg2016process,groetsch1993inverse_tl,groetsch1999inverse}.

\begin{mytcbox}[label=box:waterclocks, breakable]{Ancient water clocks}
Interestingly, ancient societies (e.g.\ ancient Greece) exploited the empirically predictable dynamics of the water level in a draining tank to measure and display the passage of time.
Specifically, the outflow \emph{clepsydra}, Greek for ``water thief'', was an open-top container with a small hole near its bottom and graduated markings on the inside. 
Filled with water then allowed to drain, the elapsed time was indicated by the liquid level with respect to the markings on the inside. \cite{bedini1962compartmented,hwang2021historical,ritner2016oriental,hejun1987research,schomberg2018karnak,mills1982newton}
The preserved Karnak clepsydra from $\sim$1300 BC \cite{schomberg2018karnak} is an inverted truncated cone. Notably, this geometry does not provide a constant rate of decrease in the water level; perhaps, though, its wider top was intended to compensate for faster outflow at higher water levels. An inverse problem pertaining to an outflow clepsydra is: what clepsydra shape provides a constant rate of decrease in the water level as it drains?
(Such a clepsydra may be obtained via a solid of revolution about the vertical axis such that the radius increases with the quartic root of the height. \cite{mills1982newton,d2021torricelli})
\end{mytcbox}


Evangelista Torricelli (1608-1647) made a fundamental observation for mathematically modeling the liquid level in a draining tank: the velocity $v$ at which liquid flows out of a small orifice in a tank is proportional to the square root of the height of liquid above the orifice, $\Delta h$, i.e. $v\propto \sqrt{\Delta h}$ \cite{mills1982newton}.
Today, we recover Torricelli's observation from Daniel Bernoulli's (1700–1782) equation \cite{welty2020fundamentals}, a mechanical energy balance applied to the steady, plug flow of an incompressible, inviscid fluid through the small orifice, neglecting frictional forces. This gives \emph{Torricelli's law}: $v=\sqrt{2 g \Delta h}$, with $g$ the acceleration due to gravity. \cite{d2021torricelli,teoman2022discharge}

A mass balance on a draining tank with Torricelli's law gives a first-order, [generally] nonlinear differential equation governing the liquid level in the tank over time \cite{groetsch1993inverse_tl,seborg2016process,debook}.
The geometry of the tank affects the dynamics of the liquid level through its cross-sectional area as a function of height.
The cross-sectional area of the orifice affects the emptying time; theoretically, it gives the volumetric flow rate out of the tank from Torricelli's law. 
However, a \emph{discharge coefficient} \cite{de2000pin,blasone2015discharge,wadhwa2021study,liu2008drainage} must be introduced into the model for agreement with experimentally-measured volumetric outflow rates \cite{farmer1992physical,driver1998torricelli,brady2009siphons,rother2024modelling,paldy1963apparatus,ivanov2014testing,williams2021vessel,pavesi2019investigating,planinvsivc2011holes,saleta2005experimental,lopac2015water,powell2012carrying}.
The discharge coefficient \cite{teoman2022discharge,hicks2014determining,blasone2015discharge,lienhard1984velocity,wadhwa2021study}
(i) is defined as the ratio of the observed outlet volumetric flow rate to that predicted by Torricelli's law and the area of the orifice \cite{hicks2014determining};
(ii) accounts for (a) the vena contracta of the liquid jet (the cross-sectional area of the liquid jet issuing from the orifice is smaller than that of the orifice, owing to fluid streamlines in the tank, near the orifice, non-perpendicular to the face of the orifice \cite{horsch2020simple}), (b) frictional losses across the orifice, and (c) non-uniformity of the velocity profile; and
(iii) depends on the rheology of the liquid, the geometry of the orifice, and, for laminar flow, the Reynolds number \cite{teoman2022discharge}. 

In contrast to the \emph{forward problem} of using this dynamic model to predict the trajectory of the liquid level in a draining tank of a given geometry, Groetsch \cite{groetsch1993inverse_tl,groetsch1999inverse} framed an \emph{inverse problem} \cite{groetsch1993inverse,neto2012introduction,tarantola2005inverse}: reconstruct the shape of the tank from its liquid level over time as it drains. 
While impossible to infer the precise geometry of the tank, one can leverage the dynamic model to determine the cross-sectional area of the tank as a function of height from the liquid level as a function of time. (The rate of decrease of the liquid level provides information about the cross-sectional area of the tank at that height. The outlet flow rate depends on the height of water only; so, wider [say, cylindrical] tanks drain more slowly.)
However, this inverse problem of reconstruction is unstable: small errors in the measured liquid level can cause large errors in the estimated area of the tank. \cite{groetsch1993inverse_tl}

Herein, we employ Bayesian statistical inversion \cite{calvetti2018inverse,waqar2023tutorial,kaipio2006statistical,dashti2013bayesian} to solve, with quantified uncertainty, an inverse problem of reconstruction: infer the shape of an exogenous, heavy solid (of unknown geometry but incapable of holding water) contained in a draining tank (of known geometry) from measurements of the liquid level in the draining tank over time.
(Because the solid displaces liquid in the tank, the rate of decrease of the liquid level provides information about the cross-sectional area of the solid at the height of the liquid.
As the tank drains, the liquid ``scans'' the area of the solid as a function of height.)
To test our ability to reconstruct the shape of the solid in the tank, we conduct tank drainage experiments with water and collect water level time series data with a level sensing strip. First, we build and calibrate (i.e., determine the posterior distribution of the parameters of) (i) a forward model of the dynamics of the water level in our draining tank and (ii) a probabilistic model of measurement noise from our level sensor.
For this task, we take length-measurements of the tank and orifice geometry and collect water level time series data from a tank drainage experiment without an exogenous solid. 
Second, we conduct a tank drainage experiment with an exogenous solid in the tank (a bottle). We then leverage this water level time series data and our calibrated forward and measurement models to obtain a posterior distribution over the area of the solid as a function of height.
Comparing with our [more] direct (via a tape measure) measurements of the solid's geometry, the inferred area of the solid traces its shape reasonably well ($<$10\% mean reconstruction error on the bottle's radius). Our approach may be practically useful to non-destructively determine the shape of an unknown solid, or the porosity, shape, and/or height of packed solid particles, inside of an opaque or underground tank.

\section{Setup for tank drainage experiments} \label{sec:expt}
\begin{figure}[h!]
\begin{center}
	\includegraphics[width=0.5\textwidth]{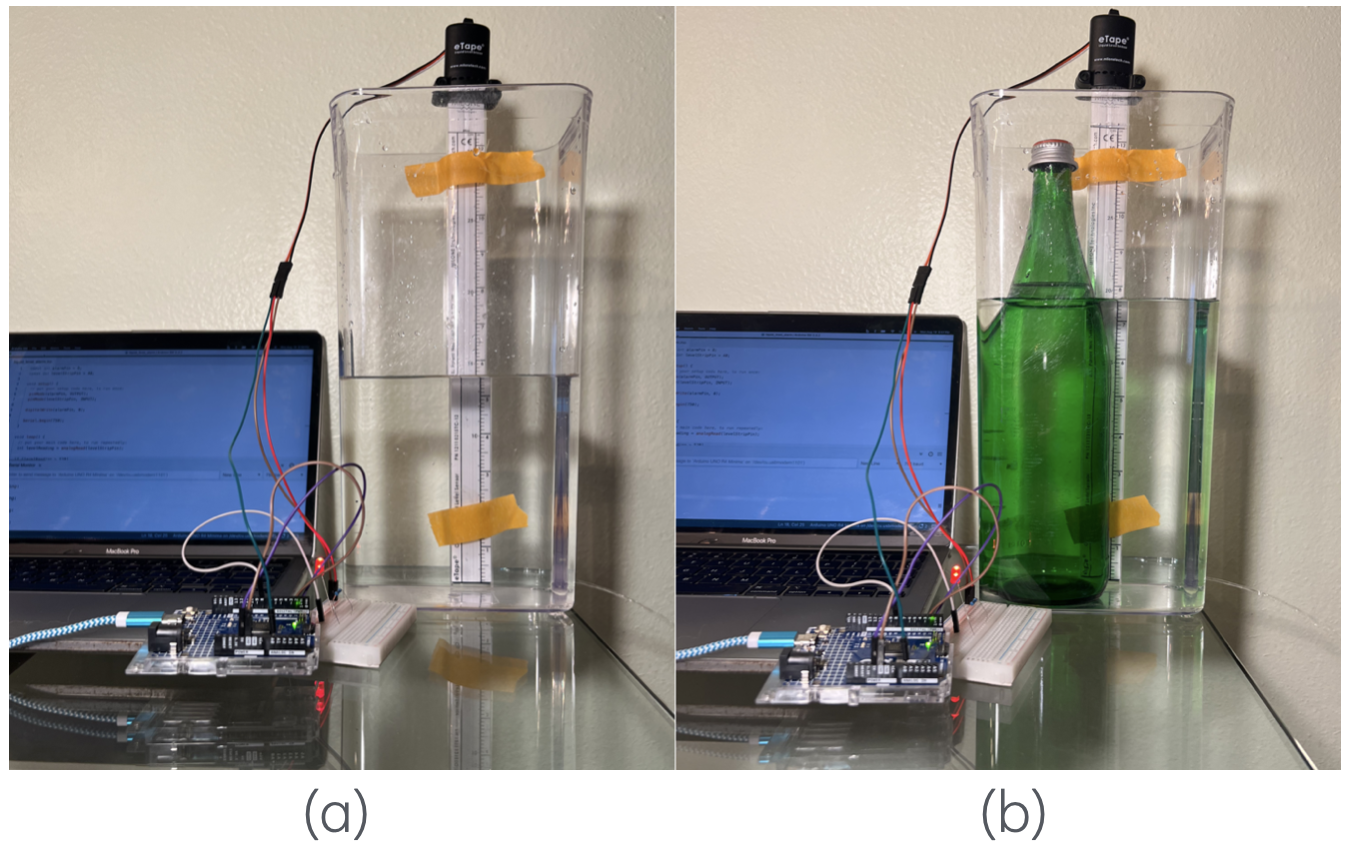}
	\caption{\textbf{Setup for tank drainage experiments} for (a) model calibration and (b) shape reconstruction.
	We fill our open-top tank (in case (b), containing an exogenous solid) with water, then allow it to drain via gravity-driven flow out of a small hole in its side, near its bottom. A liquid level strip measures the water level in the tank over time, giving time series data.
	}
	\label{fig:photo_of_tank}
\end{center}
\end{figure}

First, we describe our experimental setup in Fig.~\ref{fig:photo_of_tank} for tank drainage (of water) experiments.

We possess a plastic, open-top tank---an inverted, right, truncated cone with a rounded rectangle base (excluding the top rim). 
The cross-sectional area [parallel to the ground] of the tank as a function of height $h$ [cm] from its bottom base is:
\begin{equation}
	a(h) = \frac{h}{h_{\text{max}}}a_t + \left(1-\frac{h}{h_{\text{max}}}\right) a_b, \label{eq:a_of_h}
\end{equation}
with $h_{\text{max}}$ [cm] the height of the tank and $a_b$ and $a_t$ [cm$^2$] the area of the rounded rectangle forming the bottom and top, respectively, base of the tank.
We drilled a small, circular orifice of radius $r_o$ [cm] in the side of the tank, parallel with the ground and a small height $h_o$ [cm] (to its center) from the bottom base.

A tank draining experiment constitutes: 
(1) optionally, placing an exogenous, heavy solid inside the tank; 
(2) filling the tank with water, to an initial height $h_0$ [cm]; 
(3) at time $t=0$ [s], allowing the water to drain out (driven by gravity) through the orifice; and 
(4) collecting time series data of the water level in the tank over time, $\{(t_i, h_{\text{obs}, i}) \}$, from a liquid level sensor communicating with an Arduino microcontroller. 

The exogenous [rigid] solid (i) is heavier than water, and thus remains stationary at the bottom of the tank; (ii) displaces water in the tank; and (iii) has a geometry rendering it incapable of holding water (the solid cannot hold water iff it is locally convex at points on its (assume, smooth) surface where the normal vector to the surface is anti-parallel with the gravitational field). 
Let $\alpha(h)$ [cm$^2$] be the cross-sectional area of this solid as a function of height $h$.

(More on the liquid level sensor:)
We place an eTape\texttrademark\xspace liquid level strip vertically inside the tank, immersed in the water. 
The level strip gives an electrically-resistive output inversely proportional to the water level owing to its compression by the hydrostatic pressure of the water in which it is immersed \cite{eTape}.
The sensor communicates with an Arduino microcontroller, providing us measurements of the water level over a sequence of times. 
To map the reading from the level sensor [0-1023 integer values] to the liquid level in the tank [cm], we constructed a calibration curve (a smooth 1D spline) with a resolution of 1 cm.

\section{The forward and measurement models and likelihood} \label{sec:forward_model}
The [deterministic] \emph{forward model} predicts the water level $h(t)$ [m] in our draining tank over time $t$ [s], given the \emph{inputs} $\alpha(h)$ and $h_0$ and \emph{parameters} $a_t$, $a_b$, $h_{\text{max}}$, $r_o$, $h_o$, and the discharge coefficient $c$. 
The probabilistic \emph{measurement model} characterizes the noise contaminating a water level measurement $h_{\text{obs}}$ made by our sensor.
After we collect time series data \thedata from a tank drainage experiment, we use the forward and measurement models to construct the \emph{likelihood function}---the probability density of the data \thedata conditioned upon proposed functions/values for the inputs/parameters. 
When any of the inputs/parameters are uncertain in an inverse problem concerning the draining tank, the likelihood function quantifies the support that the data \thedata lend for each possible set of inputs/parameters.

\subsection{Forward model}
Here, we mathematically model the height of water in our draining tank as a function of time, $h(t)$. (We assume the water at the top of the tank remains flat---allowed by a slow outflow rate). 

\paragraph{Torricelli's law.}
We model the velocity $v$ [cm/s] of the jet of water flowing out of the orifice at time $t$, when the water height is $h(t)$, with Torricelli's law \cite{d2021torricelli}:
\begin{equation}
	v(h(t)) =  \sqrt{2 g(h(t)-h_o)}, \label{eq:Torricelli}
\end{equation} where $g$ [cm/s$^2$] is the acceleration due to gravity. Torricelli's law follows from Bernoulli's equation \cite{welty2020fundamentals}, a mechanical energy balance on the flow through the orifice, treating (i) the flow as steady, plug, and absent of frictional forces and (ii) the water as inviscid and incompressible.
Torricelli's law matches (i) the gain in kinetic energy, $(\Delta m) v^2/2$, when a small mass $\Delta m$ of water is ejected from the orifice, with (ii) the loss of gravitational potential energy, $(\Delta m)g(h-h_o)$, from the concomitant removal of a small slice of water of mass $\Delta m$ from the top of the body of water in the tank. \cite{groetsch1993inverse_tl,driver1998torricelli,williams2021vessel}

\paragraph{Volumetric outflow rate.}
Invoking Torricelli's law, we model the volumetric flow rate out of the tank at time $t$, when the water height is $h(t)$, as:
\begin{equation}
	c \pi r_o^2 \sqrt{2 g(h(t)-h_o)}, \label{eq:outletflow}
\end{equation}
with $c\in(0,1)$ the [dimensionless] discharge coefficient---a ``fudge factor'' to account for the vena contracta in the water jet, non-plug flow through the orifice, and frictional losses as the water flows through the orifice \cite{horsch2020simple,teoman2022discharge,hicks2014determining,blasone2015discharge,lienhard1984velocity,wadhwa2021study}. 
Note, $c=1$ corresponds with an ``ideal'' water jet with cross-sectional area equal to that of the orifice ($\pi r_o^2$), plug flow, and zero frictional losses.

\paragraph{Volume of water in the tank.} 
The volume $V$ [cm$^3$] of water in the tank at time $t$, when the water height is $h(t)$, follows from the method of cross-sections in calculus \cite{debook}:
\begin{equation}
	V(h(t))=\int_0^{h(t)} \left[a(y) - \alpha(y) \right] dy. \label{eq:volume}
\end{equation}
The integrand is the cross-sectional area of water in the tank at height $y$; subtraction of the area of the solid $\alpha(y)$ from the area of the tank $a(h)$ accounts for the solid displacing water. (We neglect the small volume of water displaced by the level sensor.) The expression assumes the solid did not retain water above $h(t)$.

\paragraph{Conservation of mass.}
Treating the water as incompressible (constant density: $\rho$ [g/cm$^3$]) and using expression~\ref{eq:outletflow} for the outflow, a mass balance on the draining tank for $t\geq 0$ gives:
\begin{equation}
	\overbrace{\diff{}{t} \Bigl( \rho V(h(t)) \Bigr )}^{\text{rate of accumulation}}= - \overbrace{\rho c \pi r_o^2 \sqrt{2 g(h(t)-h_o)}}^{\text{rate of outflow}}. \label{eq:massbalance}
\end{equation}

\paragraph{The forward model.}
Differentiating $V(h(t))$ in eqn.~\ref{eq:volume} and using the chain rule \cite{debook} to rewrite the accumulation term gives our forward model for $h(t)$:
\begin{empheq}[box={\Garybox[forward model]}]{align}
& \left[ a(h)-\alpha(h)\right] \diff{h}{t}= -c \pi r_o^2 \sqrt{2g (h(t)-h_o)}, \,\,\, t \geq 0 \label{eq:forward_model} \\
& h(0)=h_0, \nonumber
\end{empheq}
a [generally] nonlinear, first-order, ordinary differential equation (ODE) in $h(t)$ subject to an initial condition.
Given the geometry of the tank and solid through $a(h)$ and $\alpha(h)$, the discharge coefficient $c$, the orifice radius and height $r_o$ and $h_o$, and initial water height $h_0$, we numerically solve the ODE in eqn.~\ref{eq:forward_model} (with \texttt{DifferentialEquations.jl} \cite{rackauckas2017differentialequations}, which uses a Runge-Kutta method \cite{tsitouras2011runge}) to predict the water height in our draining tank over time, $h(t)$. 

\subparagraph{A regime of model invalidity.} If the radius of the orifice $r_o$ is small, surface tension of the water may stop flow out of the orifice when $h- h_o$ is positive (but small). In this regime, Torricelli's law and thus our forward model do not hold.

\subparagraph{Parameterizing the object's area as a function of height.}
We parameterize the area of the solid inside the tank as a function of height, $\alpha(h)$, with a list of $n+1$ discrete evaluations of $\alpha(h)$ on a uniform grid of points on its domain $[0, h_{\text{max}}]$:
\begin{equation}
	\boldsymbol \alpha := [\alpha_0, \alpha_1, ... \alpha_n] \label{eq:alpha}
\end{equation}
where $\alpha_i :=\alpha(i \Delta h)$ for $i \in \{0, ..., n\}$, $\Delta h = h_{\text{max}}/n$, and $0 \leq \alpha_i \leq a(i\Delta h)$ for the solid to fit in the tank.
Then, we construct the function $\alpha(h)$ for $h\in [0, h_{\text{max}}]$ via piecewise, monotonic, cubic interpolation \cite{fritsch1984method} of the function values in $\boldsymbol \alpha$. (\emph{Monotonic} interpolation prevents the unphysical outcome of $\alpha(h) > a(h)$ for any $h \in [0, h_{\text{max}}]$.) 

\subparagraph{Inputs, output, and model parameters.} 
We
consider the solid's area $\boldsymbol \alpha$ and initial water level $h_0$ as \emph{inputs} (causal factors) to the system and the water level as a function of time $h(t)$ as the \emph{output} (effect) of the system.
The \emph{model parameters} $\boldsymbol \theta \in \mathbb{R}^6$ characterize the geometry of the tank and orifice and, loosely, the rheology of water (embedded in $c$):
\begin{equation}
	\boldsymbol \theta := [h_{\text{max}}, a_b, a_t, r_o, h_o, c]. \label{eq:theta}
\end{equation}
(We omit $g$ from $\boldsymbol \theta$ because we treat it as a constant known with certainty.)
Hereafter, we write the forward model as \themodel to indicate the dependence of $h(t)$ on the inputs and parameters.

\subsection{Measurement model}
Suppose at time $t$ our water level sensor measures the height of water in the tank, providing a data point $(t, h_{\text{obs}})$ (obs for ``observation''). 
To model [unobservable] noise corrupting the measurement, we treat the measured water level $h_{\text{obs}}$ as a realization of a random variable
\begin{equation}
	H_{\text{obs}} = \themodelnomath + \Psi,
\end{equation}
with the Gaussian-distributed random variable $\Psi \overset{\text{iid}}{\sim} \mathcal{N}(0, \sigma^2)$ the noise, additive to the model output and, among multiple measurements, independent and identically-distributed (iid). 
Then, the probability density of the measured liquid level is a Gaussian centered at the model prediction:
\begin{equation}
	H_{\text{obs}} \mid h_0, \boldsymbol \alpha, \boldsymbol  \theta, \sigma^2 \overset{\text{iid}}{\sim} \mathcal{N}(\themodelnomath, \sigma^2). \label{eq:H_obs_distn}
\end{equation} We are explicit that $H_{\text{obs}}$ is conditioned upon the inputs, parameters, and noise variance and neglect model discrepancy \cite{brynjarsdottir2014learning}.


\subsection{Likelihood function}
The likelihood is the $N$-dimensional probability density of observing $\mathbf{h}_\text{obs}:=(h_{\text{obs},1}, ..., h_{\text{obs},N})$ in time series data \thedata (with each measurement time $t_i$ set) given the inputs $h_0$ and $\boldsymbol \alpha$, model parameters $\boldsymbol \theta$, and noise variance $\sigma^2$. 
Based on eqn.~\ref{eq:H_obs_distn}, the likelihood density is:
\begin{multline}
 \pi_{\text{like}}(\thedatanomath \mid h_0,\boldsymbol  \alpha, \boldsymbol \theta, \sigma^2 ) = \\ \prod_{i=1}^N \frac{1}{\sqrt{2\pi\sigma^2}} \exp \left[-\frac{1}{2}\left(\frac{h_{\text{obs}, i} - h(t_i; h_0, \boldsymbol\alpha, \boldsymbol\theta)}{\sigma} \right)^2 \right]. \label{eq:like}
\end{multline}
Once we obtain the data \thedata, we view the likelihood as a function of $h_0$, $\boldsymbol \alpha$, $\boldsymbol \theta$, and $\sigma^2$.
The likelihood function (i) scores the consistency between (a) the model predictions, with proposed values for the inputs and parameters, of the water level over time and (b) the water level time series data and, hence, (ii) quantifies the support the data lend for each possible set of values for the inputs and parameters \cite{van2021bayesian}. 

\section{Bayesian statistical inversion (BSI)} \label{sec:bsi}
Here, we outline Bayesian statistical inversion (BSI)  \cite{calvetti2018inverse,waqar2023tutorial,kaipio2006statistical,dashti2013bayesian,allmaras2013estimating}, a tool for solving inverse problems while incorporating prior information and quantifying uncertainty. 
Later, we employ BSI to, using water level time series data from tank drainage experiments,  
(1, parameter inference) calibrate our forward and measurement models and 
(2, reconstruction) infer the shape of a solid in the tank.

Under BSI, we treat the uncertain input(s) and/or parameters of a tank drainage experiment as random variables and model their probability distributions.
The probability density over input/parameter space expresses our knowledge and beliefs about the inputs/parameters, namely 
(i) they most likely lie in the region containing the bulk of the density and (ii) the spread (concentration) of the density quantifies our uncertainty (certainty) about them. 


The BSI approach to inverse problems with draining tanks follows three stages:

\paragraph{Specify the prior.}
Before we allow the tank to drain and observe water level time series data, we express our knowledge and [to an extent, subjective] beliefs about the inputs and parameters through a \emph{prior} probability density $\pi_{\text{pr}}(h_0, \boldsymbol \alpha, \boldsymbol \theta, \sigma^2)$.
The [marginal] prior density on an input/parameter can range from 
(i) diffuse (e.g.\ a uniform distribution) if we lack knowledge about it, adopting the principle of indifference, to 
(ii) informative (e.g.\ a Gaussian with a small variance) if we possess a [noisy] measurement of it. 
\cite{van2021bayesian}

\paragraph{Gather data and construct the likelihood.}
Next, we conduct a tank drainage experiment (perhaps, with the tank containing an exogenous solid) to collect water level time series data, \thedata. 
When considered against the forward and measurement models, these data provide new information about the inputs and parameters. The likelihood function $\pi_{\text{like}}(\thedatanomath \mid h_0,\boldsymbol  \alpha, \boldsymbol \theta, \sigma^2 )$ quantifies the support the data lend for different inputs and parameters.

\paragraph{Update the prior to a posterior.}
In light of the data \thedata, we invoke Bayes's theorem  \cite{van2021bayesian,calvetti2018inverse} to update our prior density of the inputs and parameters to a \emph{posterior} density
\begin{multline}
	\pi_{\text{post}}(h_0, \boldsymbol \alpha, \boldsymbol \theta, \sigma^2 \mid \thedatanomath) \propto \\ 
	\pi_{\text{like}}(\thedatanomath \mid h_0,  \boldsymbol \alpha, \boldsymbol \theta, \sigma^2 ) 
	\pi_{\text{pr}}(h_0, \boldsymbol\alpha, \boldsymbol \theta, \sigma^2).
	 \label{eq:post}
\end{multline} 
A compromise between the likelihood and prior,
the posterior density expresses our knowledge and beliefs about the inputs and parameters \emph{conditioned upon the data} [and, implicitly, the structure of our forward and measurement models]. 
Thus, the posterior represents the raw, uncertainty-quantifying solution to the inverse problem.
Note, uncertainty (entropy) remains in the posterior because
 (i) our measurements are noisy and 
 (ii) perhaps, even noiseless data cannot fully constrain a subset of the parameters/inputs.
(The normalizing factor for the posterior, the \emph{evidence}, is a high-dimensional integral that can, in principle, be computed from the likelihood and prior.)


To approximate the posterior, we conduct a Markov chain Monte Carlo (MCMC) simulation \cite{robert1999monte,van2021bayesian} that sequentially draws a sample of inputs/parameters $(h_0, \boldsymbol \alpha, \boldsymbol \theta, \sigma^2 )$ from the posterior density $\pi_{\text{post}}(h_0, \boldsymbol \alpha, \boldsymbol \theta, \sigma^2 \mid \thedatanomath)$. 
From many samples from the posterior, we draw empirical posterior distributions, compute credible intervals, and plot samples of liquid level trajectories when paired with the forward model.
Specifically, we use the No-U-Turn Sampler (NUTS) \cite{hoffman2014no} implemented in the probabilistic programming language \cite{gordon2014probabilistic} \texttt{Turing.jl} \cite{ge2018turing}.
A key advantage of MCMC samplers is that we can circumvent computing the evidence.

\section{Results}
We now employ BSI to, using water level time series data from tank drainage experiments,  
(1, parameter inference) calibrate our forward and measurement models then
(2, reconstruction) exploit our calibrated model to infer the shape of a solid contained in the tank.

\subsection{Phase I: Bayesian calibration of the forward and measurement models}
\label{sec:phaseI}
Our objective now is to calibrate then test our forward and measurement models.
At this point, the model parameters $\boldsymbol \Theta$ and measurement noise variance $\Sigma^2$ are highly uncertain and, thus, treated as random variables (notation: upper-case $\boldsymbol \Theta$ denotes the random variable, the lower-case version $\boldsymbol \theta$ denotes a realization).
To gather information about $\boldsymbol \Theta$ and $\Sigma^2$, we (a) take length-measurements of the tank and orifice geometry and (b) collect water level time series data from a tank drainage experiment devoid of a solid (certainly, $\boldsymbol \alpha = \mathbf{0}$).
The \emph{calibrated model} constitutes eqns.~\ref{eq:forward_model} and \ref{eq:H_obs_distn} with the resulting posterior distribution of $\boldsymbol \Theta$ and $\Sigma^2$. We test the calibrated model for its capability to predict the liquid level trajectory in a replicate tank drainage experiment.


\subsubsection{Experimental setup}
We set up a tank draining experiment with an initial water level (measured by the level strip) $h_{0, \text{obs}}=26.54$\,cm. 
With certainty, the tank does not contain an exogenous solid. See Fig.~\ref{fig:naked_tank}.

\subsubsection{Prior distributions} 
We express our knowledge and beliefs about the inputs and parameters, before allowing the tank to drain and observing water level time series data, by constructing a prior density $\pi_{\text{pr}}(h_0, \boldsymbol \alpha, \boldsymbol \theta, \sigma^2)$ variable-by-variable (assuming independence). Together with the forward model in eqn.~\ref{eq:forward_model}, $\pi_{\text{pr}}(h_0, \boldsymbol \alpha, \boldsymbol \theta, \sigma^2)$ gives a prior distribution over water level trajectories in the tank, \themodel. 
Fig~\ref{fig:prior_train} shows samples of functions \themodel from the prior, which encodes a broad range of possible trajectories of the liquid level. 

\subparagraph{Tank geometry.} We use a measuring tape to make length-measurements of the dimensions of the tank.
Specifically, we measure the length, width, and perimeter of the rounded rectangle forming the top and bottom of the tank,
$l_{t, \text{obs}}=14.6$\,cm, $w_{t, \text{obs}}=9.0$\,cm, $p_{t, \text{obs}}=44.3$\,cm,
$l_{b, \text{obs}}=13.4$\,cm, $w_{b, \text{obs}}=7.8$\,cm, and $p_{b, \text{obs}}=40.1$\,cm, and the height of the tank $h_{\text{max}, \text{obs}}=28.6$\,cm.
Informed by these [noisy and imprecise] measurements, we impose Gaussian prior distributions:
\begin{align}
L_{t,b} &\sim \mathcal{N}(l_{t,b, \text{obs}}, \sigma_\ell^2) \\
W_{t,b} &\sim \mathcal{N}(w_{t,b, \text{obs}}, \sigma_\ell^2) \\
P_{t,b} &\sim \mathcal{N}(p_{t,b, \text{obs}}, \sigma_\ell^2) \\
H_{\text{max}} &\sim \mathcal{N}(h_{\text{max}, \text{obs}}, \sigma_\ell^2)
\end{align}
where $\sigma_\ell=0.1$\,cm is the [assumed] standard deviation of our length-measurements, based on the markings on our measurement tape. 
For the area of the rounded rectangle forming the top and bottom of the tank,
we (i) solve for the radius of the circles at the four corners $R_{t/b}$ consistent with these measurements using $P_{t,b}=2(\pi R_{t,b} + L_{t,b}+W_{t,b}-4R_{t,b})$ then (ii) compute the area \cite{rounded_rect}
\begin{multline}
	A_{t,b}= (L_{t,b}-2R_{t,b})(W_{t,b}-2R_{t,b}) \\ + 2R_{t,b} (L_{t,b}+W_{t,b} -4R_{t,b}) + \pi R_{t,b}^2.
\end{multline}
The prior distributions of $A_t$ and $A_b$ follow accordingly. 
(This gives point estimates $a_{t, \text{obs}}=129.9$\,cm and $a_{b, \text{obs}}=103.0$\,cm.)

\subparagraph{Orifice height and geometry.} 
We measure the height of the orifice in the side of the tank as $h_{o, \text{obs}}=0.9$\,cm and impose an informative prior:
\begin{equation}
H_o \sim \mathcal{N}(h_{o, \text{obs}}, \sigma_\ell^2).
\end{equation}
Informed by the manufacturer-reported diameter (5/64\,in) of the drill bit we used to drill the hole in the tank, we estimate the radius of the orifice as $r_{o, \text{obs}}=0.1$\,cm and impose the prior:
\begin{equation}
R_o \sim \mathcal{N}(r_{o, \text{obs}}, \sigma_d^2), \label{eq:R_o_prior}
\end{equation}
where $\sigma_d= 0.001$\,cm expresses uncertainty in the bit radius and its translation into an orifice.

\subparagraph{Exogenous solid geometry.}
With certainty, no solid resides in the tank, so $\boldsymbol \alpha=\mathbf{0}$. Technically, this is a Dirac delta prior density on $\mathbf{A}$, $\pi_{\text{pr}}(\boldsymbol \alpha)=\delta(\boldsymbol \alpha - \mathbf{0})$.


\subparagraph{Discharge coefficient.} 
Our prior distribution on the discharge coefficient is weakly informed:
\begin{equation}
	C \sim \mathcal{N}(0.65, 0.25^2).
\end{equation} The mean is a reported discharge coefficient for water flow through a round orifice \cite{hicks2014determining}.

\subparagraph{Variance of measurement noise.} 
We impose a uniform prior on the standard deviation of the noise emanating from our water level sensor:
\begin{equation}
\Sigma \sim \mathcal{U}(0\,\text{cm}, 0.5\,\text{cm}),
\end{equation} whose generous upper bound is informed by our experience in calibrating the liquid level strip.

\subparagraph{Initial water level.} We impose an informative prior distribution on the initial water level based on the initial reading of the water level sensor:
\begin{equation}
	H_0 \mid \sigma \sim \mathcal{N}(h_{0, \text{obs}}, \sigma^2).
\end{equation}

\subsubsection{Water level time series data from a tank-drainage experiment} At time $t:=0$, we allow the tank to drain of water while our level sensor records the water level over time. 
Fig.~\ref{fig:posterior_train} shows the resulting water level time series data \thedata.

\subsubsection{Posterior distribution}
The posterior distribution $\pi_{\text{post}}(h_0, \boldsymbol \alpha, \boldsymbol \theta, \sigma^2 \mid \thedatanomath)$ follows from our prior and likelihood function constructed from the data \thedata (see eqn.~\ref{eq:post}). 
We employ NUTS to draw $2000$/chain samples of inputs/parameters from the posterior over three independent chains, excluding the first half of each chain discarded for ``burn-in''. 
(When constructing our likelihood, we omit the last three data points in the time series \thedata because surface tension stopped flow out of the orifice despite $h>h_o$---a phenomenon not accounted for in the model.)

\begin{figure*}[!ht]
    \centering
        \begin{subfigure}[b]{0.2\textwidth}
    	\includegraphics[width=\textwidth]{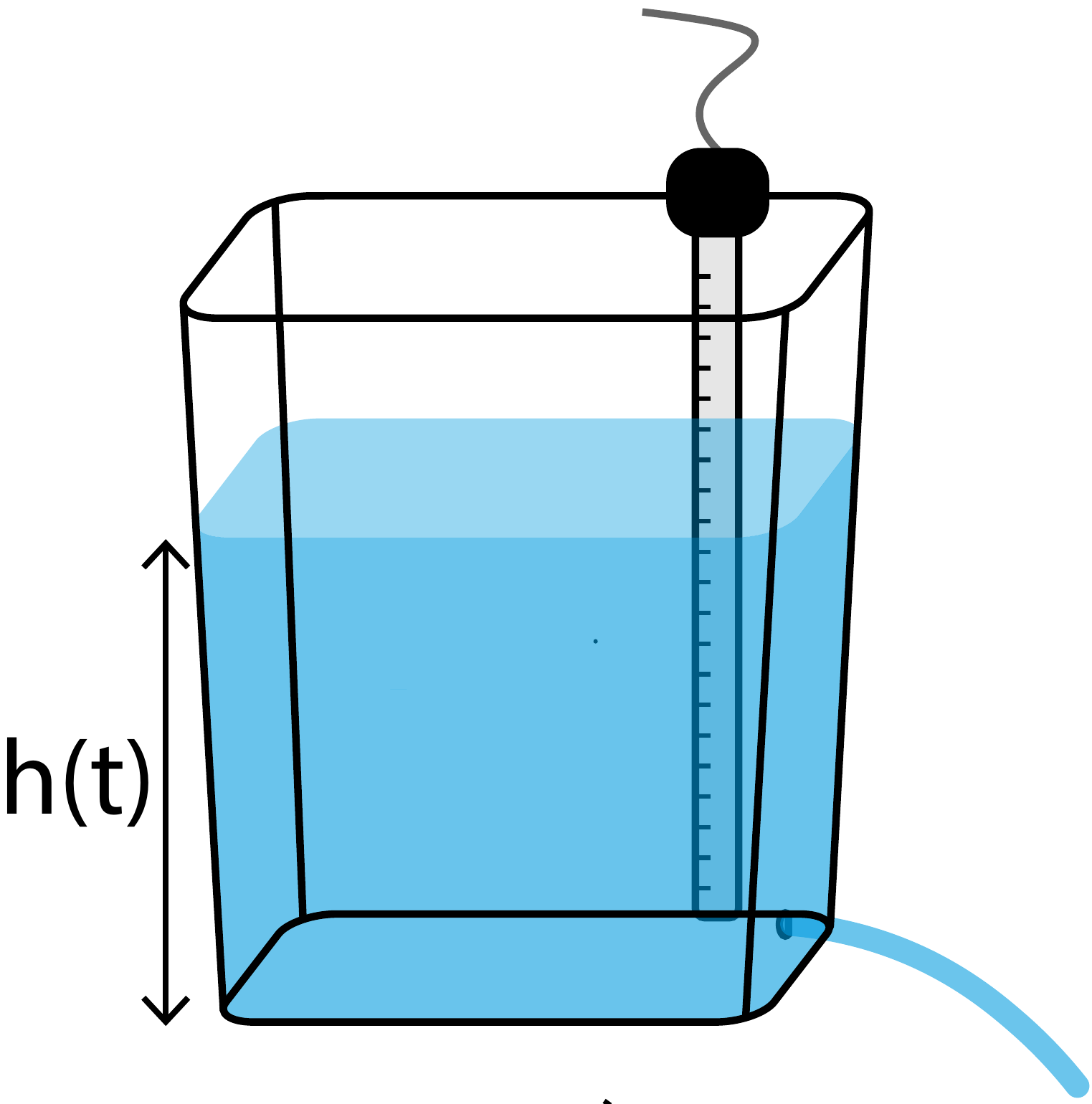}
	\caption{Experimental setup} \label{fig:naked_tank}
    \end{subfigure}
    
     \begin{subfigure}[b]{0.35\textwidth}
    	\includegraphics[width=\textwidth]{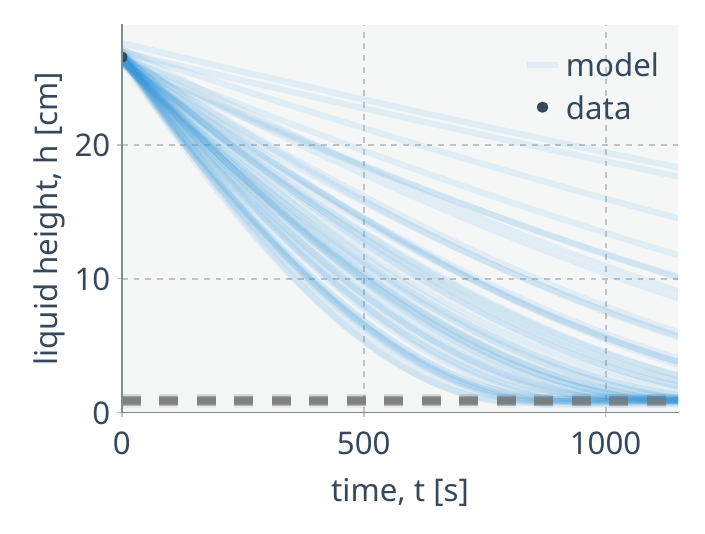}
	\caption{Prior distribution} \label{fig:prior_train}
    \end{subfigure}
     \begin{subfigure}[b]{0.35\textwidth}
    	\includegraphics[width=\textwidth]{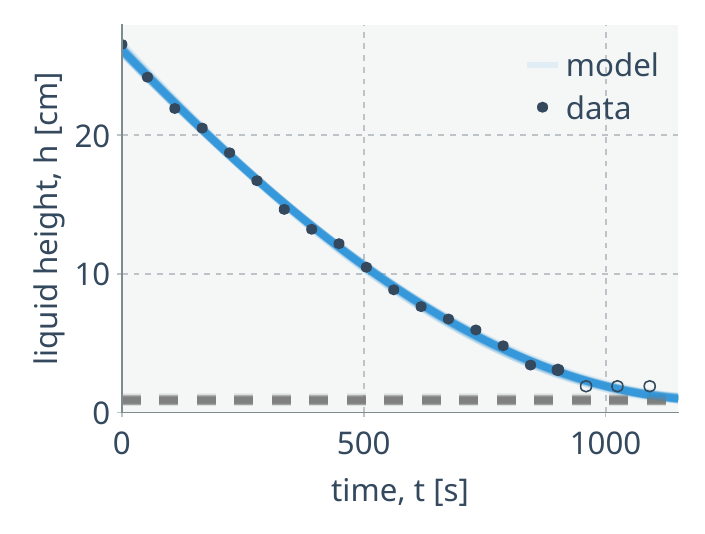}
	\caption{Data \& posterior distribution} \label{fig:posterior_train}
    \end{subfigure}
    
     \begin{subfigure}[b]{\textwidth}
     \center
    	\includegraphics[width=0.65\textwidth]{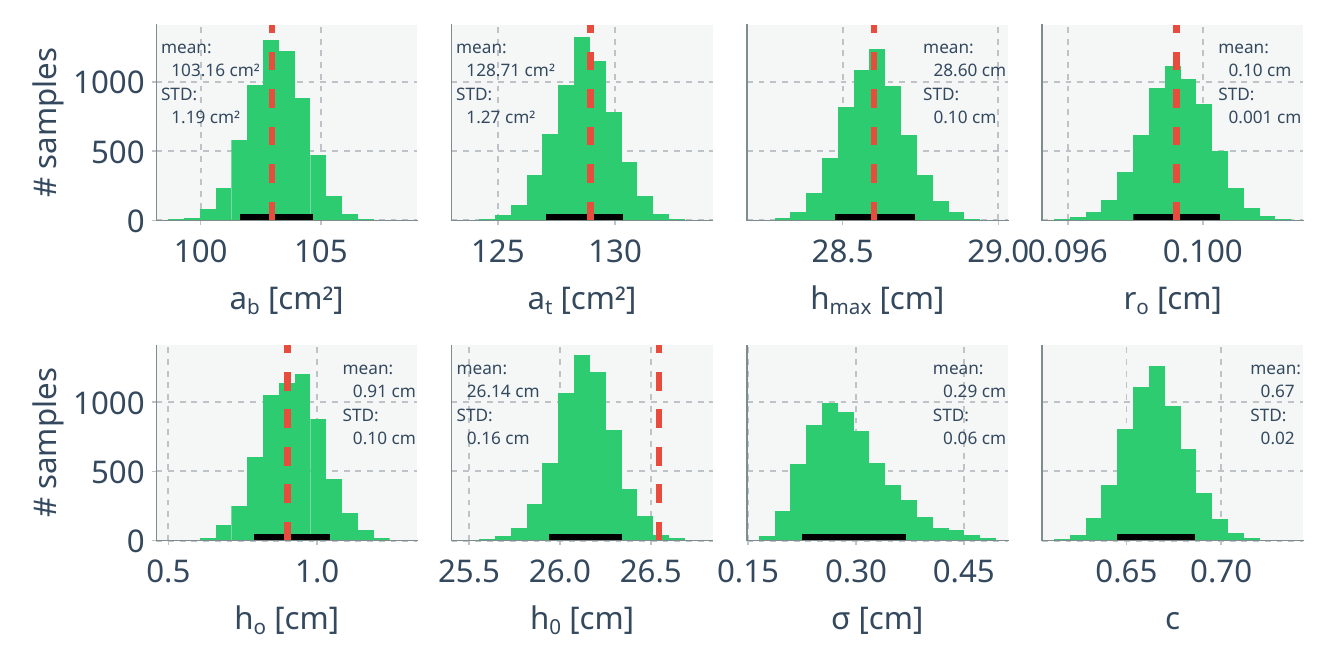}
	\includegraphics[width=0.34\textwidth]{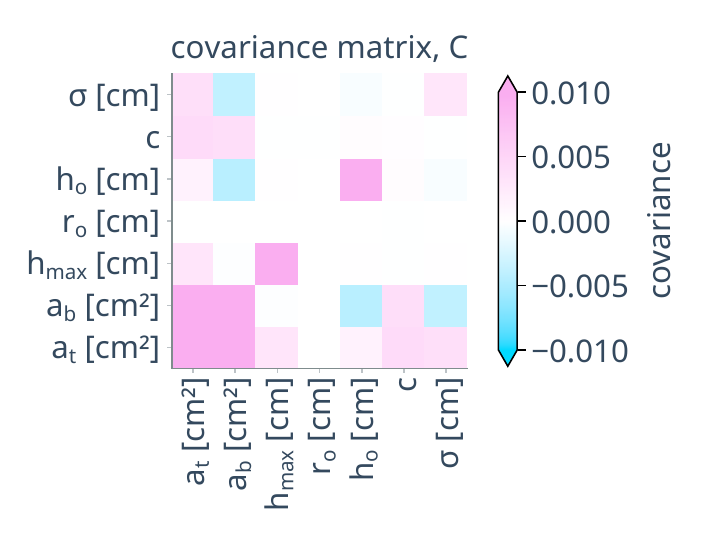}
	\caption{Posterior distribution} \label{fig:posterior_train_theta}
    \end{subfigure}
    \caption{
      \textbf{Model calibration.}
      (a) We fill our tank (devoid of an exogenous solid) with water, then, at $t=0$, allow it to drain through the orifice in its side. A level sensor measures the water level.     
      (b) Samples of water level trajectories from the prior.
      (c) Time series data from the tank drainage experiment and samples of water level trajectories from the posterior (hollow points omitted).
       (d) Visualizing the posterior. Left: marginals. Right: covariance matrix.      
      }
\end{figure*}


Fig.~\ref{fig:posterior_train_theta} visualizes the empirical posterior distribution. 
The top panel shows the [marginal] empirical posterior distributions (histograms) of the parameters and initial water level along with their 80\% equal-tailed credible intervals. 
For the variables we measured with our measuring tape, the credible intervals are centered at the measurements. 
The measured initial water level, however, falls outside of the credible interval; a higher initial water level than measured is more consistent with the time series data.
The bottom panel shows the covariance matrix of the parameters and input. E.g. $c$ positively covaries with both $a_t$ and $a_b$ because the dynamics of $h(t)$ depend on the ratio $c/a(h)$.


For a posterior predictive check, Fig.~\ref{fig:posterior_train} shows water level trajectories \themodel, with $(\thevars)$ sampled from the posterior distribution, agree reasonably with the measured water levels over time (mean absolute residual: 0.26\,cm). 
Comparing the prior and posterior distributions of water level trajectories in Fig.~\ref{fig:prior_train} and Fig.~\ref{fig:posterior_train}, the reduced variance in the trajectories reflects information about the parameter vector gained from considering the water level time series data against the forward and measurement models.

The forward and measurement models in eqn.~\ref{eq:forward_model} and \ref{eq:H_obs_distn} together with the posterior distribution $\pi_{\text{post}}(\boldsymbol \theta, \boldsymbol \sigma^2 \mid \thedatanomath)$ (with $h_0$ marginalized out and $\boldsymbol \alpha=\mathbf{0}$) constitute the \emph{calibrated model}.
We approximate the posterior as a multi-variate Gaussian
\begin{equation}
	\begin{bmatrix} \boldsymbol \Theta \\ \Sigma \end{bmatrix} \mid \thedatanomath \sim \mathcal{N}(\mathbf{m}, \mathbf{C}) \label{eq:post_theta_sigma}
\end{equation}
with mean $\mathbf{m}$ and covariance matrix $\mathbf{C}$ computed from our MCMC samples from the posterior. 
(See Fig.~\ref{fig:posterior_train_theta} for the means of each parameter in $\mathbf{m}$ and the covariance matrix $\mathbf{C}$.)

\subsubsection{Testing the calibrated model with a replicate experiment}
\begin{figure}[h!]
    \centering
    	\includegraphics[width=0.4\textwidth]{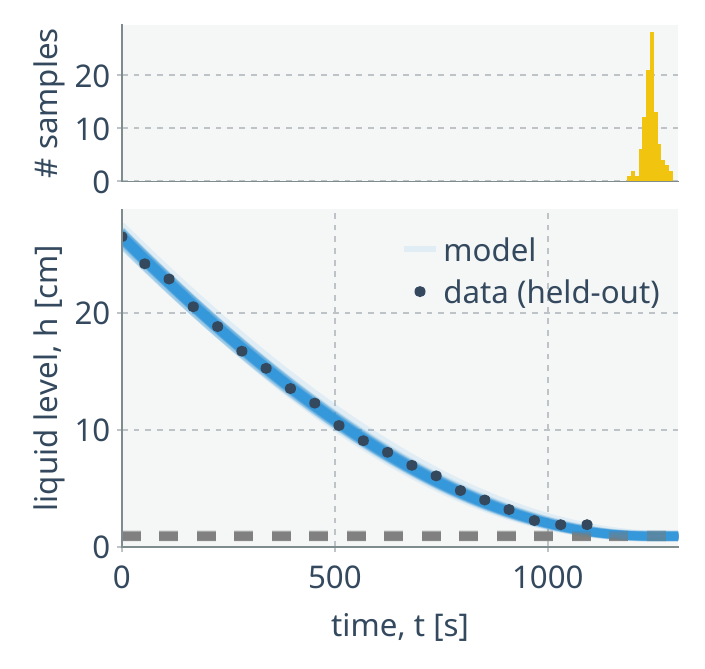}
    \caption{
      \textbf{Testing the calibrated forward model.}
      (Bottom) Comparison between water level time series data from a replicate tank drainage experiment and water level trajectories predicted by the calibrated forward model. 
      (Top) Distribution of predicted emptying times.
      } \label{fig:test}
\end{figure}

Finally, we test our calibrated model for its capability to predict the dynamics of the water level in our draining tank in a new, replicate experiment (also, without an exogenous solid, so $\boldsymbol \alpha=\mathbf{0}$) with measured initial water level $h_{0, \text{obs}}=26.49$\,cm. 
We sample water level trajectories \themodel predicted by the calibrated model by (1) sampling a $\boldsymbol \theta$ and $\sigma$ from the posterior distribution in eqn.~\ref{eq:post_theta_sigma}, (2) sampling an initial water level from $H_0 \mid \sigma \sim \mathcal{N}(h_{0, \text{obs}}, \sigma^2)$, then (3) numerically solving the forward model to obtain \themodel.
Fig.~\ref{fig:test} shows the water level time series data collected from the replicate experiment, \thedata, agree well with the trajectories \themodel predicted by the calibrated model (mean absolute residual: 0.28\,cm). 
(For kicks, we also show the distribution of predicted tank-emptying times, $T_e := \min_{t \geq 0 :\, h(t)=h_o} t$.)

\subsection{Phase II: Bayesian inference of the shape of a solid inside the tank} \label{sec:phaseII}
Finally, we exploit our calibrated model to infer the shape of an exogenous, stationary, heavy solid (incapable of holding water itself) inside of our tank from water level time series data as it drains.
Our objective is to obtain the posterior distribution of $\mathbf{A}$ (random variable version of $\boldsymbol \alpha$) characterizing the cross-sectional area of the solid as a function of height, $\alpha(h; \mathbf{A})$.

\begin{figure*}[ht]
    \centering
        \begin{subfigure}[b]{0.2\textwidth}
    	\includegraphics[width=\textwidth]{tank_w_bottle.pdf}
	\caption{Experimental setup} \label{fig:tank_w_bottle}
    \end{subfigure}
     \begin{subfigure}[b]{0.8\textwidth}
    	\includegraphics[width=0.48\textwidth]{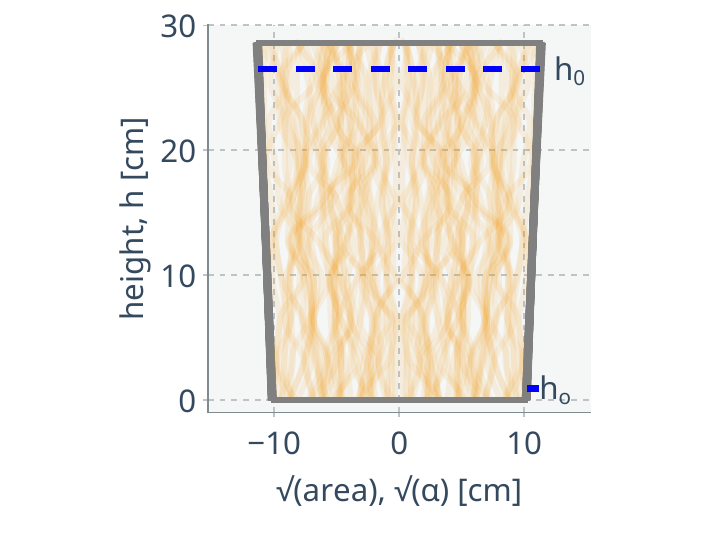}
	\includegraphics[width=0.48\textwidth]{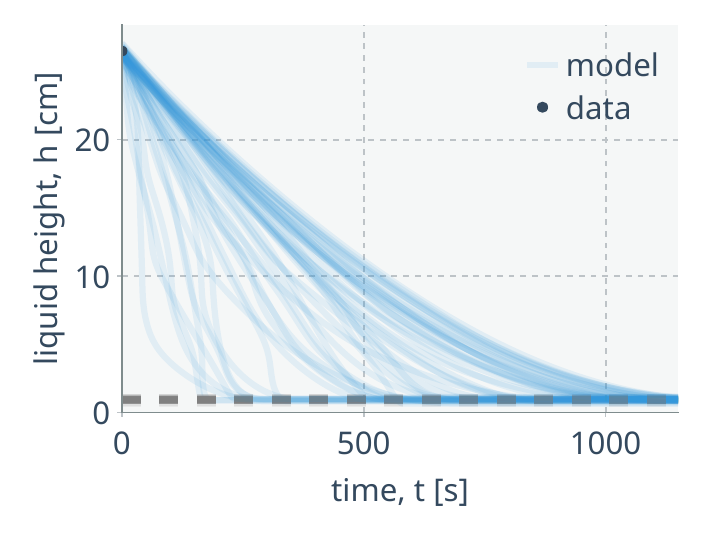}
	\caption{Prior distribution} \label{fig:prior_area}
    \end{subfigure}
    
     \begin{subfigure}[b]{0.4\textwidth}
    	\includegraphics[width=\textwidth]{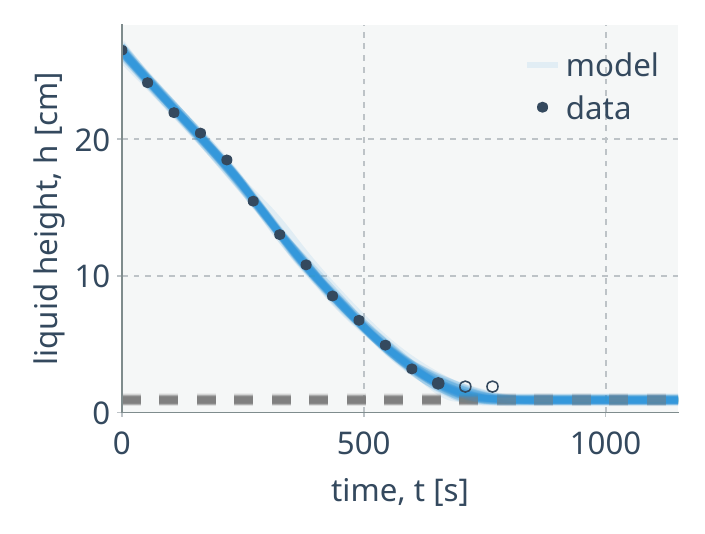}
	\caption{Data \& posterior distribution} \label{fig:posterior_object}
    \end{subfigure}
    \begin{subfigure}[b]{0.44\textwidth}
    	\includegraphics[width=\textwidth]{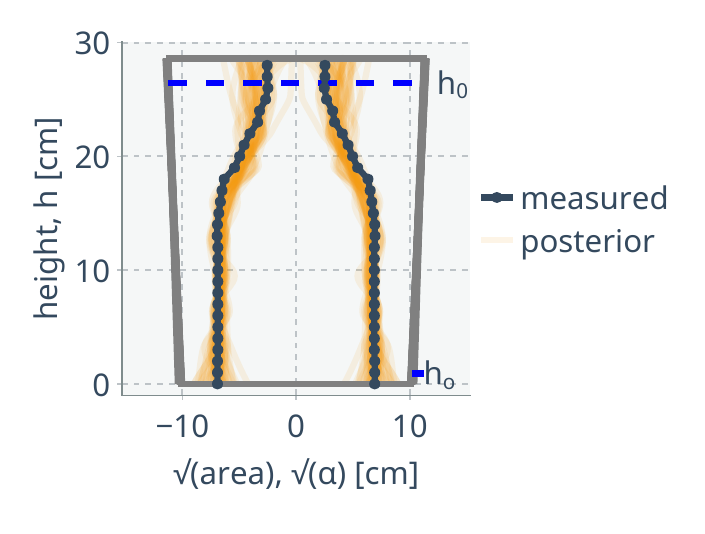}
	\caption{Posterior distribution} \label{fig:posterior_area}
    \end{subfigure}
    \caption{
      \textbf{Reconstructing the shape of an exogenous solid in the draining tank.} 
      (a) We fill our tank, containing a heavy glass bottle, with water, then, at $t=0$, allow it to drain through the orifice in its side.
      (b) Samples from the prior of (left) the square root of the solid's area as a function of height (orange lines), along with the square root of the area of the tank (gray) and (right) associated liquid level trajectories.
      (c) Time series data from the tank drainage experiment and samples of water level trajectories from the posterior.
      (d) Samples of the square root of the solid's inferred area as a function of height from the posterior distribution (orange lines) compared with held-out, direct measurements (points) and the tank's area (gray).
      }
\end{figure*}

\subsubsection{Experimental setup}
We set up a tank drainage experiment by:
(i) placing a solid object---a large, capped, glass bottle filled with water [so it sinks]---inside of the tank, which remains stationary;
(ii) filling the tank to an initial water level (measured by the level strip) of $h_{0, \text{obs}}=26.49$\,cm. 
See Fig.~\ref{fig:tank_w_bottle}.

\subsubsection{Prior distributions}
Before allowing the tank to drain, we express our prior knowledge and beliefs about the inputs and parameters below.
Fig.~\ref{fig:prior_area} displays samples of our prior distribution over (left) the solid's area as a function of height (for perspective, the effective tank boundary is also shown) and (right) associated liquid level trajectories. 
Our prior over the solid's area encodes a uniform distribution over the effective average radius of the solid in the tank, but promotes a smoothly changing area as a function of height. 

\subparagraph{The parameter vector and measurement noise variance.}
Adopting the adage ``yesterday's posterior is today's prior'' \cite{calvetti2010subjective}, 
our multi-variate Gaussian posterior distribution of the parameter vector $\boldsymbol \Theta$ and noise variance $\Sigma^2$ in eqn.~\ref{eq:post_theta_sigma} from model calibration now serves as our prior distribution.
Thereby, we exploit our \emph{calibrated} model for this reconstruction problem.
(We assume the solid in the tank does not affect the discharge coefficient.)

\subparagraph{Initial water level.} We again impose an informative prior distribution on the initial water level based on the initial reading of the water level sensor:
\begin{equation}
	H_0 \mid \sigma \sim \mathcal{N}(h_{0, \text{obs}}, \sigma^2).
\end{equation}

\subparagraph{Shape of the exogenous solid.}
We adopt the principle of indifference and impose a diffuse prior over the size of the solid in the tank, but impose smoothness on the cross-sectional area of the solid as a function of height to counter instability\cite{groetsch1993inverse_tl}. 
By imposing a diffuse prior on the solid's size, we allow the water level time series data to ``speak for itself'' about the solid's size.
Specifically, we impose a uniform distribution on the square root of the area of the bottom base of the solid, $A_0$ (the random variable version of $\alpha(0)$):
\begin{equation}
	\sqrt{A_0} \mid a_b \sim \mathcal{U}(0, \sqrt{a_b}).
\end{equation}
So, we believe 
(i) the area of the solid's bottom base can span from zero (corresponding to the lack of an exogenous solid) to the area of tank's bottom base $a_b$ (corresponding to the largest solid that can fit in the tank) and
(ii) the effective radius of the solid (hence, we take the square root of $A_i$) is uniformly distributed.
For the smoothness prior \cite{calvetti2018inverse}, we sequentially correlate the solid's effective radius at the sequence of heights: 
\begin{equation}
 \sqrt{A_i} - \sqrt{A_{i-1}} \sim \mathcal{N}(0, \gamma^2) \text{ for } i \in \{1, ..., n\}.
\end{equation} 
The zero mean Gaussian distribution on the difference in the effective radius of the solid at two adjacent heights means we expect the function $\sqrt{\alpha(h)}$ to be locally flat and change slowly. 
The variance hyperparameter $\gamma^2$ controls the degree to which we promote smoothness in $\sqrt{\alpha (h)}$ (smaller $\gamma^2$ promotes more smoothness). 
Finally, for physicality, we also truncate each random variable $A_i$ so it falls within zero to the area of the tank at that height.


\subsubsection{Water level time series data from a tank-drainage experiment}
Next, at time $t=0$, we allow our tank containing the exogenous solid to drain of water. Fig.~\ref{fig:posterior_object} shows the time series data of the water level \thedata obtained from our level sensor. Compared to the water level dynamics in the calibration experiment where the tank was devoid of an exogenous solid (Fig.~\ref{fig:posterior_train}), the tank drained more quickly, owing to the displacement of water by the solid in the tank.

\subsubsection{Posterior distribution}
Again, we employ NUTS to obtain samples from the posterior $\pi_{\text{post}}(h_0, \boldsymbol \alpha, \boldsymbol \theta, \sigma^2 \mid \thedatanomath)$. 

Fig.~\ref{fig:posterior_area} shows posterior samples of the square root of the area of the solid as a function of height---the BSI solution to the reconstruction problem. 
For comparison, we also show (i) the bottle's [roughly-] ground-truth area at different heights, obtained from our length-measurements of its circumference at different heights, and (ii) for perspective, posterior samples of the tank's area as a function of height. First, the posterior of the solid's shape exhibits much lower variance than the prior in Fig.~\ref{fig:prior_area}---reflecting the information the water level time series data provided about the solid's shape. 
Second, the posterior of the solid's shape agrees reasonably well with the measured area of the solid (mean residual radius: 0.35\,cm; mean bottle radius 3.2\,cm), but a little bias is present in $h\in[15, 20]$\,cm.
Third, note the variance in the posterior over the solid's area in $h<h_o$ and $h>h_0$ is larger than in $h_o < h < h_0$. This reflects the impossibility for the water level data to provide information about the solid's area where the water was unable to ``scan'' the solid. 
For $h<h_o$ and $h>h_0$, the smoothness prior restricts the variance of the predicted area---essentially, extrapolating. 

As a posterior predictive check, Fig.~\ref{fig:posterior_object} displays samples of liquid level trajectories \themodel from the posterior distribution, which match the water level time series data \thedata reasonably well (mean absolute residual: 0.35\,cm). 

\subsubsection{Conclusion} Overall, Fig.~\ref{fig:posterior_area} illustrates our ability to infer the shape of a solid inside a draining tank with quantified uncertainty---using (i) a calibrated dynamic model of the water level in the tank and (ii) water level time series data over the process of draining. 

\section{Conclusions and Discussion}
We demonstrated our ability to reconstruct, with quantified uncertainty, the cross-sectional area of a heavy, exogenous solid (incapable of holding water) inside a tank from (i) a calibrated dynamic model of the liquid level in the tank as it drains (driven by gravity) through a small orifice and (ii) measurements of the liquid level in the tank over time as it drains.
First, we constructed, Bayesian-calibrated, and tested a forward model for the dynamics of the liquid level in our tank draining of water.
The calibrated model predicted the liquid level trajectory in our tank in a replicate experiment with a mean absolute residual of 0.28\,cm.
Then, we exploited our Bayesian-calibrated forward model to reconstruct the shape of an exogenous, heavy solid (a bottle) inside our draining tank using water level time series data. 
Essentially, the water ``scanned'' the area of the solid as a function of its height as the tank drained. 
The posterior distribution of the bottle's area as a function of height agreed reasonably well with our [more] direct measurements of its area, with $\sim 10$\% reconstruction error over the bottle's radius. 

Our methodology herein could be employed throughout engineering and the applied sciences to non-destructively characterize the shape of an exogenous solid inside of an opaque or inaccessible (e.g., underground \cite{gephart2010short}) liquid-holding tank.
With some modification, one could also infer the height and/or porosity of a bed of small, heavy solid particles (e.g. gravel, sand) packed into a tank \cite{guellouz2020estimation}; we could infer the \emph{shape} of the particles as well, by coupling with a packing model \cite{zhang2006relationship}. 
While our methodology requires access to the liquid level in the tank as it drains, we could infer the liquid level from the volumetric flow rate or the geometry of the liquid jet exiting from the tank \cite{groetsch1999inverse}.

\paragraph{Can we infer the precise geometry of a solid in an opaque, draining tank?}
Our ambition to reconstruct the geometry of an exogenous solid inside a draining tank from its liquid level over time was limited to seeking the area of the solid a function of height; determining the precise geometry of the solid is impossible because the dynamics of $h(t)$ depend solely on the area of the solid as a function of height (see eqn.~\ref{eq:forward_model})---and, there are multiple non-congruent solids with an identical area as a function of height (e.g., a cylinder and rectangular prism with the same cross-sectional area as a function of height). We speculate that, by conducting multiple tank-draining experiments with the tank tilted and held at different angles, we may be able to reconstruct the precise geometry of the solid in the tank. The key idea is that two non-congruent solids with an identical area as a function of height may exhibit different areas as a function of height when tilted; thus, they can be distinguished by tilting the tank. A complication is that gravity could cause the shape to rotate when the tank is tilted.

\paragraph{Remarks.} (1) We could have calibrated our forward and measurement models using data from a tank drainage experiment where the tank \emph{did} contain an exogenous solid, provided the area of the solid as a function of height were known with some certainty. (2) If we knew the solid's shape belonged to a certain category (e.g.\ a right circular cone), we could have parameterized its area as a function of height with fewer variables (e.g.\ height and base radius). 

\paragraph{Extensions.}
(1) Our forward model of the dynamics of the liquid level in an \emph{open-top} tank, containing a \emph{solid}, draining of liquid through a \emph{small} orifice in its side \emph{without} additional input/output of liquid could be extended to handle inverse problems concerning
(a) closed tanks, where the air pressure in the headspace is not constant and atmospheric during draining; 
(b) larger orifices, where Torricelli's law does not hold and perhaps the liquid at the top does not remain flat (necessitating modeling of flow streamlines as in Refs.~\cite{mathew2014numerical,sakri2017numerical});
(c) additional inputs/outputs of liquid by adding source/sink terms; 
(d) Mariotte's bottle \cite{kirevs2006mariotte}; and/or
(e) a flexible exogenous object compressible by hydrostatic pressure, such as a balloon \cite{muller2004rubber}. 
(2) During Bayesian model calibration, a model discrepancy \cite{brynjarsdottir2014learning,kennedy2001bayesian} function could capture any bias---i.e., a difference between the true dynamics of the liquid level in the tank and the model $h(t; h_0^*,  \boldsymbol \alpha^*, \boldsymbol \theta^*)$ with the true/best-fit (indicated by $*$) initial water level, solid geometry, and parameters.
(3) Finally, another interesting inverse problem pertaining to a draining tank is time reversal: infer the initial water level from measurement(s) of the water level at later times.

\paragraph{Other geometric inverse problems.}
Examples of other geometric inverse problems, where the unknown is a geometric shape \cite{ameur2004level,burger2001level,harbrecht2013numerical,kawakami2020stabilities}, include inferring
(i) from boundary measurements: the shape and location of cracks in a solid \cite{nishimura1991boundary}, fouling on the inner wall of a pipe \cite{chen2011inverse}, mines underground \cite{delbary2007inverse,lopez2003detection}, scattering objects like submarines \cite{yaman2013survey} underwater \cite{buchanan2004marine}, and inclusions or cavities inside thermal conductors \cite{wang2018numerical,nakamura2015reconstruction};
(ii) the shape of obstacles in blood vessels from flow measurements \cite{aguayo2021distributed,nolte2022inverse}; 
(iii) the shape of a membrane from the sound it produces when it vibrates \cite{kac1966can}; 
and (iv) the potential energy of the interaction between a particle and a surface as a function of distance from the distribution of [fluctuating] distance measurements \cite{prieve1999measurement}.
Geometric inverse problems leading to optimal engineering designs include determining: 
(i) the region in input space, constituting design and operating parameters, for a chemical reactor that gives an output falling in a desired set \cite{alves2023inverse,gazzaneo2019process}; 
(ii) the shape of (a) an adhesive pillar that gives a desired interfacial stress distribution \cite{kim2020designing};
(b) particles giving a desired packing stiffness \cite{miskin2013adapting}, self-assembly behavior \cite{sacanna2013engineering}, or optical properties \cite{forestiere2016inverse};
(c) an airfoil giving desired aerodynamic properties \cite{sun2015artificial};
(d) the blades of a heat exchanger achieve a desired heat exchange \cite{hilbert2006multi};
or
(e) a fixed-bed chemical reactor giving a desired conversion \cite{courtais2021shape}. 

\section*{Data and code availability} The raw liquid level time series data and Julia code to reproduce our results are available at \url{github.com/SimonEnsemble/tank_inverse_problem}.

\enlargethispage{20pt}

\begin{acknowledgments}
CMS acknowledges support from the Intel Corporation through their Mindshare Program - Broadening Participation in Science and Engineering Higher Education. 
Thanks to 
Adrian Henle for feedback on our manuscript and Axel Rodriguez for help articulating the condition for a solid to be incapable of holding water.
\end{acknowledgments}

\vskip2pc

\bibliographystyle{RS} 

\bibliography{refs} 

\end{document}